\documentclass[12pt,a4paper,english]{article}
\usepackage[T1]{fontenc}
\usepackage[latin9]{inputenc}
\usepackage{babel}
\usepackage{amsmath}
\usepackage{amssymb}
\usepackage[unicode=true,
 bookmarks=true,bookmarksnumbered=false,bookmarksopen=false,
 breaklinks=false,pdfborder={0 0 1},colorlinks=false]
 {hyperref}
\hypersetup{pdftitle={Numerical Polynomial Homotopy Continuation Method and String Vacua},
 pdfauthor={Dhagash Mehta},
 pdfsubject={Algebraic Geometry, Mathematical Methods in Physics},
 pdfstartview={FitH},bookmarksopen,bookmarksnumbered,bookmarksopenlevel=2}
\usepackage{breakurl}

\makeatletter

\usepackage{a4}\usepackage[all]{xy}
\usepackage{multirow}

\usepackage{cite}

\usepackage{babel}

\makeatother

\begin{document}
\vspace*{-1.5cm}

\vspace{1cm}

\begin{center}
{\LARGE Numerical Polynomial Homotopy Continuation Method and String
Vacua }
\par\end{center}

\vspace{0.75cm}

\begin{center}
Dhagash Mehta%
\footnote{Email: dbmehta@syr.edu%
}
\par\end{center}

\vspace{0.15cm}

\begin{center}
Physics Department, Syracuse University, Syracuse, NY 13244, USA.
\par\end{center}

\vspace{1cm}

\begin{abstract}
Finding vacua for the four dimensional effective theories for supergravity
which descend from flux compactifications and analyzing them according
to their stability is one of the central problems in string phenomenology.
Except for some simple toy models, it is, however, difficult to find
all the vacua analytically. Recently developed algorithmic methods
based on symbolic computer algebra can be of great help in the more
realistic models. However, they suffer from serious algorithmic complexities
and are limited to small system sizes. In this article, we review
a numerical method called the numerical polynomial homotopy continuation
(NPHC) method, first used in the areas of lattice field theories,
which by construction finds \textit{all }of the vacua of a given potential
that is known to have only isolated solutions. The NPHC method is
known to suffer from no major algorithmic complexities and is \textit{embarrassingly
parallelizable}, and hence its applicability goes way beyond the existing
symbolic methods. We first solve a simple toy model as a warm up example
to demonstrate the NPHC method at work and compare the results with
the available results from the symbolic methods. We then show that
all the vacua of a more complicated model of M theory compactified
on the coset $\frac{SU(3)\times U(1)}{U(1)\times U(1)}$, which has
an $SU(3)$ structure, can be obtained by the NPHC method using a
desktop machine in just about one hour, a feat which was reported
to be prohibitively difficult by the existing symbolic methods. Finally,
we compare the various technicalities between the two methods.
\end{abstract}
\clearpage{}

\newpage{}



\section{Introduction}

A lot of current research in string phenomenology is focused on developing
methods to find and analyze vacua of four dimensional effective theories
for supergravity descended from flux compactifications. Stated in
explicit terms, one is interested in finding all the vacua (usually,
isolated stationary points) of the scalar potential $V$ of such a
theory. In particular, given a Kähler potential $K$, and a superpotential
$W$, for uncharged moduli fields, the scalar potential is given by

\[
V=e^{K}[\mathit{\mathit{\mathit{K^{A\bar{B}}}}}\ D_{A}W\ D_{\bar{B}}\bar{W}-3|W|^{2}]
\]
 where, $D_{A}$ is the Kähler derivative $\partial_{A}+\partial_{A}K$
and $\mathit{K}^{A\bar{B}}$ is the inverse of $\mathit{K_{A\bar{B}}}=\partial_{A}\partial_{\bar{B}}K$.
Once the vacua are found, one can then classify them by either using
the eigenvalues of the Hessian matrix of $V$ or by introducing further
constraints such as $W=0$.

Finding all the stationary points of a given potential $V$, amounts
to solving the stationary equations, i.e., solving the system of equations
consisting of the first derivatives of $V$, with respect to all the
fields, equated to zero. The stationary equations for $V$ arising
in the string phenomenological models are usually nonlinear. In the
perturbative limit, $W$ usually has a polynomial form. This is an
important observation since we can then use the algebraic geometry
concepts and methods to extract a lot of information about $V$. Solving
systems of nonlinear equations is usually a highly non-trivial task.
However, if the system of stationary equations has polynomial-like
non-linearity, then the symbolic methods based on the Gröbner basis
technique can be used to solve the system \cite{Gray:2009fy}. These
symbolic methods ensure that all the stationary points are obtained
when the computation finishes. Roughly speaking, for a given system
of multivariate polynomial equations, a set of which is called an
ideal, the so-called Buchberger Algorithm (BA) or its refined variants
can compute a new system of equations, called a Gröbner basis \cite{CLO:07}.
For the systems known to have only isolated solutions, called $0$-dimensional
ideals, a Gröbner basis always has at least one univariate equation
and the subsequent equations consist of increasing number of variables,
i.e., it is in a \textit{triangular form}%
\footnote{Note that this is only true for a few specific types of monomial orderings.
For other monomial ordering, the new system of equations may not have
a triangular form.%
}. The solutions of a Gröbner basis is always the same as the original
system, but the former is easier to solve due to its triangular form
as the univariate equation can be solved either analytically or numerically
quite straightforwardly. Then by back-substituting the solutions in
the subsequent equations and continually solving them we can find
all the solutions of the system%
\footnote{Using the Gröbner basis methods, one can also deal with systems which
have at least one free variable, called positive dimensional ideals.
However, in this review we only focus on the $0$-dimensional ideals.%
}. It should be noted that the BA reduces to Gaussian elimination in
the case of linear equations, i.e., it is a generalization of the
latter. Similarly it is also a generalization of the Euclidean algorithm
for the computation of the Greatest Common Divisors of a univariate
polynomial. Recently, more efficient variants of the BA have been
developed to obtain a Gröbner basis, e.g., F4 \cite{Faugere99anew},
F5 \cite{Faug:02} and Involution Algorithms~\cite{2005math1111G}.
Symbolic computation packages such as Mathematica, Maple, Reduce,
etc., have built-in commands to calculate a Gröbner basis. Singular
\cite{DGPS}, COCOA \cite{CocoaSystem} and MacCaulay2 \cite{M2}
are specialized packages for Gröbner basis and Computational Algebraic
Geometry, available as freeware. MAGMA \cite{BCP:97} is also such
a specialized package available commercially.

In \cite{Gray:2006gn,Gray:2008zs,Gray:2007yq}, it was shown that
one does not need to solve the system using the Gröbner basis techniques,
in the usual sense, in order to extract some of the important information
such as the dimensionality of the ideal, the number of real roots
in the system etc. but one can indirectly obtain this information
by computing the so-called primary decomposition of the ideal (still
using the Gröbner basis technique internally). This was a remarkable
success as it allowed one to work on non-trivial models and extract
a lot of information using a regular desktop machine only. The authors
of these papers also made a very helpful computational package, called
Stringvacua \cite{Gray:2008zs}, publicly available. Stringvacua is
a Mathematica interface to Singular and has string phenomenology specific
utilities which makes the package quite useful to the users.

However, even with such tricks, there are a few problems with the
symbolic methods: the BA is known to suffer from exponential space
complexity, i.e., the memory (Random Access Memory) required by the
machine blows up exponentially with the number of variables, equations,
terms in each polynomial, etc. So even for small sized systems, one
may not be able to compute a Gröbner basis, nor the related objects
such as primary decomposition of the ideal. It is also usually less
efficient for systems with irrational coefficients. Another drawback
is that the BA is highly sequential, i.e., very difficult to efficiently
parallelize.

Below we explain a novel numerical method, called the numerical polynomial
homotopy continuation (NPHC) method, which overcomes all the shortcomings
of the Gröbner basis methods. The method, first introduced in particle
physics and condensed matter theory areas in Refs. \cite{Mehta:2009,2009iwqg.confE..25M,Mehta:2011xs},
where all the stationary points of a multivariate function called
the lattice Landau gauge fixing functional\cite{vonSmekal:2007ns,vonSmekal:2008es,Mehta2011,Kastner:2011zz}
were found using the NPHC method. Below, we begin by describing the
NPHC method for the univariate case before generalizing it to the
multivariate case. We then consider a toy model that is used in the
Stringvacua manual, and also a compactified M theory model. Finding
all the vacua using the symbolic methods for both these models is
already known to be prohibitively difficult. We briefly describe the
models and explain how the corresponding stationary equations can
be viewed as having polynomial form. With the help of the NPHC method,
we find all the isolated vacua for the model and give a technical
comparison between both the symbolic and numerical methods. After
mentioning a few other important aspects of the NPHC method in the
Frequently Asked Questions section, we conclude the paper.

\section{The Numerical Polynomial Homotopy Continuation Method}

Here, we explain the numerical polynomial homotopy continuation method.
Let us begin by exemplifying the method for the univariate case.

Firstly, we know that for a single variable equation, $f(x)=\sum_{i=0}^{k}a_{i}x^{i}$,
with coefficients $a_{i}$ and the variable $x$ both defined over
$\mathbb{C}$, the number of solutions is exactly $k$ if $a_{k}\neq0$,
counting multiplicities. This powerful result comes from the Fundamental
Theorem of Algebra. To get all roots of such single variable polynomials,
there exist many numerical methods such as the \textit{companion matrix}
trick for low degree polynomials and the divide-and-conquer techniques
for high degree polynomials. Here we present the Numerical Polynomial
Homotopy Continuation (NPHC) by first describing it for the univariate
case which can then be extended to the multivariate case in a straightforward
manner. We follow Refs.~\cite{SW:95,Li:2003} throughout this section
unless specified otherwise.

The strategy behind the NPHC method is: first write down the equation
or system of equations to be solved in a more general parametric form,
solve this system at a point in parameter space where its solutions
can be easily found, and finally tracking these solutions from this
point in parameter space to the point in parameter space corresponding
to the original system/problem. This approach can be applied to many
types of equations (e.g., non-algebraic equations) which exhibit a
continuous dependence of the solutions on the parameters, but there
exist many difficulties in making this method a primary candidate
method to solve a set of non-algebraic equations. However, for reasons
that will be clear below, this method works exceptionally well for
polynomial equations.

To clarify how the method works, we first take a univariate polynomial,
say $z^{2}-5=0$, pretending that we do not know its solutions (i.e.,
$z=\pm\sqrt{5}$). We then begin by defining the more general parametric
family
\begin{equation}
H(z,t)=(1-t)(z^{2}-1)+t(z^{2}-5)=z^{2}-(1+4t)=0\label{eq:one_var_homotopy}
\end{equation}
 where $t\in[0,1]$ is a parameter. For $t=0$, we have $z^{2}-1=0$
and at $t=1$ we recover our original problem. The problem of getting
all solutions of the original problem now reduces to tracking solutions
of $H(z,t)=0$ from $t=0$ where we know the solutions, i.e., $z=\pm1$,
to $t=1$. The choice of $z^{2}-1$ in Eq.~(\ref{eq:one_var_homotopy}),
called the \textit{start system}, should be clear now: this system
has the same number of solutions as the original problem and is easy
to solve. For multivariate systems, a clever choice of a start system
is essential in reducing the computation, and the discussion about
this issue will follow soon. Here, we briefly mention the numerical
methods used in path-tracking from $t=0$ to $t=1$. One of the ways
to track the paths is to solve the differential equation that is satisfied
along all solution paths, say $z_{i}^{*}(t)$ for the $i^{th}$ solution
path,
\begin{equation}
\frac{dH(z_{i}^{*}(t),t)}{dt}=\frac{\partial H(z_{i}^{*}(t),t)}{\partial z}\frac{dz_{i}^{*}(t)}{dt}+\frac{\partial H(z_{i}^{*}(t),t)}{\partial t}=0.
\end{equation}
 This equation is called the Davidenko differential equation. Inserting
(\ref{eq:one_var_homotopy}) in this equation, we have
\begin{equation}
\frac{dz_{i}^{*}(t)}{dt}=-\frac{2}{z_{i}^{*}(t)}.
\end{equation}
 We can solve this initial value problem numerically (again, pretending
that an exact solution is hitherto unknown) with the initial conditions
as $z_{1}^{*}(0)=1$ and $z_{2}^{*}(0)=-1$. The other approach is
to use Euler's predictor and Newton's corrector methods. This approach
works well too. We do not intend to discuss the actual path tracker
algorithm used in practice, but it is important to mention that in
these path tracker algorithms, almost all apparent difficulties have
been resolved, such as tracking singular solutions, multiple roots,
solutions at infinity, etc. It is also important to mention here that
in the actual path tracker algorithms the homotopy is randomly complexified
to avoid singularities, i.e., taking
\begin{equation}
H(z,t)=\gamma(1-t)(z^{2}-1)+t(z^{2}-5)=0,
\end{equation}
 where $\gamma=e^{i\theta}$ with $\theta\in\mathbb{R}$ chosen randomly.

It is shown that for a generic value of the complex $\gamma$ the
paths are well-behaved for $t\in[0,1)$, i.e., for the whole path
except the end-point. This makes sure that there is no singularity
or bifurcation along the paths. This is a remarkable trick, called
the $\gamma$-trick, since this is the reason why we can claim that
the NPHC method is guaranteed to find \textit{all} solutions. Note
that $\gamma=1$, for example, is not a generic value.

There are several sophisticated numerical packages well-equipped with
path trackers such as Bertini \cite{BHSW06}, PHCpack~\cite{Ver:99},
PHoM~\cite{GKKTFM:04}, HOMPACK~\cite{MSW:89} and HOM4PS2 \cite{GLW:05,Li:03}.
They all are available freely from their respective research groups.

In the above example, the PHCpack with its default settings gives
the solutions
\begin{eqnarray*}
z & = & \pm2.23606797749979\pm i\,0.00000000000000.
\end{eqnarray*}
 Thus, it gives the expected two solutions of the system with a very
high numerical precision.

\subsection{Multivariate Polynomial Homotopy Continuation}

We can now generalize the NPHC method to find all the solutions of
a system of multivariate polynomial equations, say $P(x)=0$, where
$P(x)=(p_{1}(x),\dots,p_{m}(x))$ and $x=(x_{1},\dots,x_{m})$, that
is \textit{known to have isolated solutions} (i.e., a $0$-dimensional
ideal). To do so, we first need to have some knowledge about the expected
number of solutions of the system. There is a classical result, called
the \textit{Classical Bï¿œzout Theorem}, that asserts that for a system
of $m$ polynomial equations in $m$ variables the maximum number
of solutions in $\mathbb{C}^{m}$ is $\prod_{i=1}^{m}d_{i}$, where
$d_{i}$ is the degree of the $i$th polynomial. This bound, called
the Classical Bï¿œzout Bound (CBB), is exact for generic values (i.e.,
roughly speaking, non-zero random values) of coefficients. The \textit{genericity}
is well-defined and the interested reader is referred to Ref.~\cite{SW:95}
for details.

Based on the CBB, we can construct a \textit{homotopy}, or a set of
problems, similar to the aforementioned one-dimensional case, as
\begin{equation}
H(x,t)=\gamma(1-t)Q(x)+t\; P(x)=0,
\end{equation}
 where $Q(x)$ is a system of polynomial equations, $Q(x)=(q_{1}(x),\dots,q_{m}(x))$
with the following properties:
\begin{enumerate}
\item The solutions of $Q(x)=H(x,0)=0$ are known or can be easily obtained.
$Q(x)$ is called the \textit{start system} and the solutions are
called the \textit{start solutions}.
\item The number of solutions of $Q(x)=H(x,0)=0$ is equal to the CBB for
$P(x)=0$.
\item The solution set of $H(x,t)=0$ for $0\le t\le1$ consists of a finite
number of smooth paths, called homotopy paths, each parametrized by
$t\in[0,1)$.
\item Every isolated solution of $H(x,1)=P(x)=0$ can be reached by some
path originating at a solution of $H(x,0)=Q(x)=0$.
\end{enumerate}
We can then track all of the paths corresponding to each solution
of $Q(x)=0$ from $t=0$ to $t=1$ and reach $P(x)=0=H(x,1)$. By
implementing an efficient path tracker algorithm, we can get all the
isolated solutions of a system of multivariate polynomials just as
in the univariate case.

\noindent The homotopy constructed using the CBB is called the \textit{Total
Degree Homotopy}. The start system $Q(x)=0$ can be taken, for example,
as
\begin{equation}
Q(x)=\left(\begin{array}{c}
x_{1}^{d_{1}}-1\\
x_{2}^{d_{2}}-1\\
.\\
.\\
.\\
x_{m}^{d_{m}}-1
\end{array}\right)=0,\label{eq:Total_Degree_Homotopy}
\end{equation}
 where $d_{i}$ is the degree of the $i^{th}$ polynomial of the original
system $P(x)=0$. Eq.~(\ref{eq:Total_Degree_Homotopy}) can be easily
solved and its total number of solutions (the start solutions) is
$\prod_{i=1}^{m}d_{i}$, all of which are non-singular. The Total
Degree Homotopy is a very effective and popular homotopy whose variants
are used in the actual path trackers.

For the multivariate case, a solution is a set of numerical values
of the variables which satisfies each of the equations within a given
tolerance, $\triangle_{\mbox{sol}}$ ($\sim10^{-10}$ in our set up).
Since the variables are allowed to take complex values, all the solutions
come with real and imaginary parts. A solution is a real solution
if the imaginary part of each of the variables is less than or equal
to a given tolerance, $\triangle_{\mathbb{R}}$ ($\sim10^{-7}$ is
a suitable choice for the equations we will be dealing with in the
next section, below which the number of real solutions does not change).
All of these solutions can be further refined to an \textit{arbitrary
precision} limited by the machine precision.

The obvious question at this stage would be if the number of real
solutions depends on $\triangle_{\mathbb{R}}$. To resolve this issue,
we use a recently developed algorithm called alphaCertified which
is based on the so-called Smale's $\alpha$-theory \cite{2010arXiv1011.1091H}.
This algorithm certifies the real non-singular solutions of polynomial
systems using both exact rational arithmetic and arbitrary precision
floating point arithmetic. This is a remarkable step, because using
alphaCertified we can prove that a solution classified as a real solution
is actually a real solution independent of $\triangle_{\mathbb{R}}$,
and hence these solutions are as good as the \textit{exact solutions}.

\section{A Toy Model}

Here, we apply the NPHC method to a toy model from the examples given
in the Stringvacua package. The Kähler potential for this model is
given as

\[
K=-3\log(T+\bar{T}),
\]
 and the superpotential is given as

\[
W=a+bT^{8}.
\]
 Here, $a$ and $b$ are parameters. Note that the field $T$ comes
along with its complex conjugate. So even though they can be treated
as different variables by merely relabeling them, they are not actually
independent variables. To avoid this problem, we can write them in
terms of real and imaginary parts, i.e., $T=t+i\,\tau$ with $\tau$,
and $t$ are real. Finally, we get the potential as

\begin{eqnarray*}
V & = & \frac{1}{3t}(4b(5b(t^{2}+\tau^{2})^{7}-3a(t^{6}-21t^{4}\tau^{2}+35t^{2}\tau^{4}-7\tau^{6}))),
\end{eqnarray*}
 which has $2$ variables. To find the stationary points of $V$,
we need to solve the system of equations consisting of the first order
derivatives of $V$, with respect to both variables $t$ and $\tau$,
equated to zero, i.e.,

\begin{align*}
\frac{\partial V}{\partial t} & =\frac{1}{3t^{2}}(4b(5b(13t^{2}-\tau^{2})(t^{2}+\tau^{2})^{6}-3a(5t^{6}-63t^{4}\tau^{2}+35t^{2}\tau^{4}+7\tau^{6})))=0,\\
\frac{\partial V}{\partial\tau} & =\frac{1}{3t}(56b\tau(5b(t^{2}+\tau^{2})^{6}+a(9t^{4}-30t^{2}\tau^{2}+9\tau^{4})))=0.
\end{align*}

We also note that the stationary equations in this example involve
denominators. Since we are not interested in the solutions for which
the denominators are zero, we clear them out by multiplying them with
the numerators appropriately.

Using the symbolic methods, this task is known to be difficult for
general numerical (i.e., floating points) values of parameters $a$
and $b$, with the computation continuing indefinitely \cite{Gray:2006gn,Gray:2009fy}.

Firstly, we used the Stringvacua package to compute the dimension
of the ideal which turned out to be $0$ for generic values of $a$
and $b$, i.e., the system of equations has only isolated solutions.
Note that to actually find the solutions of the system, we have to
put some numerical values for $a$ and $b$. The Gröbner basis techniques,
as mentioned above, work much better for the cases where parameters
are rational. We first use the same values, $a=1$ and $b=1$, as
used in the Stringvacua manual. Then, we use the command {}``NumRoots''
which computes the number of real roots of the system, i.e., $7$
in this case, in less than a minute on a desktop machine.

Let us now turn our attention to solving this system using the NPHC
method. Firstly, the CBB for this system is $182$. We used both Bertini
and HOM4PS2 to track all these paths. Both took around one minute
to solve this system: there are $86$ complex (including real) finite
solutions, out of which $36$ solutions are real. Out of the $36$
real solutions, six of them are distinct solutions (multiplicity one)
and the only other distinct solution $(t,\tau)=(0,0)$ which comes
with multiplicity $30$. Thus, there are $7$ distinct solutions as
expected from the Stringvacua's {}``NumRoots'' command. However,
we should mention that the Stringvacua package does not give any information
about the multiplicity of the solutions, as seen in this example,
whereas the NPHC method gives all the solutions with its multiplicities
making the method already useful for this simple example. Not only
that, but the NPHC also gives the infinite solutions (which are the
solutions on the projective space but not on the affine space): the
running example has $2$ infinite solutions both coming with multiplicity
$\mbox{48}$. Thus, the total number of solutions in this case, $50+6+(1\times30)+(2\times48)$,
is indeed the same as the CBB.

Note that in these equations all the denominators were multiples of
$t$. The condition that none of the denominators is zero can be imposed
algebraically by adding a constraint equation as $1-z\, t=0$ with
$z$ being an additional variable. Thus there are now $3$ equations
in $3$ variables. Note that in the Stringvacua package the denominators
are thrown away by multiplying each equation appropriately, but the
additional equation is not included in the final ideal. In the package,
one can of course use the {}``Saturation'' command in order to ensure
that this equation is properly taken into account.

We can again solve the above system $3$ equations in $3$ variables
using the Bertini and HOM4PS2. The CBB of this new system is $364$.
In the end, there are $56$ finite complex solutions out of which
there are six real solutions, all with multiplicity $1$. There are
no infinite solutions in this case. This should be expected since
the only multiple real solution in the previous system was when the
denominator was zero. After adding the constraint equation, we have
got rid of this solution and hence left with the rest of the six distinct
solutions. Finally, the real solutions (throwing the very small imaginary
parts out) are:

\begin{align*}
\{t,\tau\}= & \{\{-0.5204819146691344,0.7148265478403096\},\\
 & \{0.5204819146691421,-0.7148265478403003\},\\
 & \{-0.5204819146691322,-0.7148265478403191\},\\
 & \{0.520481914669129,0.7148265478403104\},\\
 & \{0.8325249117100803,0\},\{-0.8325249117100793,0\}\}.
\end{align*}

Since we have all the real solutions, we can now compute the Hessian
of $V$ at these solutions and separate out the physically interesting
vacua. Since the purpose of this article is to introduce the NPHC
method only, we refrain from discussing the interesting physics of
these solutions here. A detailed analysis of these solutions and the
solutions of other systems will be published elsewhere. For now we
discuss how the two methods, the symbolic algebra methods and the
NPHC, compare with each other.

\section{A Model of Compactified M Theory}

Here, we take an example of M theory compactified on the coset $\frac{SU(3)\times U(1)}{U(1)\times U(1)}$
from Ref. \cite{Micu:2006ey} which is also considered in Ref. \cite{Gray:2006gn}.
The coset has $SU(3)$ structure. The corresponding Kähler and superpotential
are

\begin{eqnarray*}
K & = & -4\log(-i(U-\bar{U}))-\log(-i(T_{1}-\bar{T}_{1})(T_{2}-\bar{T}_{2})(T_{3}-\bar{T}_{3})),\\
W & = & \frac{1}{\sqrt{8}}(4U(T_{1}+T_{2}+T_{3})+2T_{2}T_{3}-T_{1}T_{3}-T_{1}T_{2}+200).
\end{eqnarray*}

Here, we use $T_{i}=-it_{i}+\tau_{i}$, for $i=1,2,3$, and $U=-ix+y$.
Then the potential is

\begin{eqnarray*}
V & = & \frac{1}{256t_{1}t_{2}t_{3}x^{4}}(40000+t_{3}^{2}\tau_{1}^{2}-400\tau_{1}\tau_{2}-4t_{3}^{2}\tau_{1}\tau_{2}+4t_{3}^{2}\tau_{2}^{2}+\tau_{1}^{2}\tau_{2}^{2}-400\tau_{1}\tau_{3}\\
 &  & +800\tau_{2}\tau_{3}+2\tau_{1}^{2}\tau_{2}\tau_{3}-4\tau_{1}\tau_{2}^{2}\tau_{3}+\tau_{1}^{2}\tau_{3}^{2}-4\tau_{1}\tau_{2}\tau_{3}^{2}+4\tau_{2}^{2}\tau_{3}^{2}-24t_{2}t_{3}x^{2}\\
 &  & +4t_{3}^{2}x^{2}-24t_{1}(t_{2}+t_{3})x^{2}+4\tau_{1}^{2}x^{2}+8\tau_{1}\tau_{2}x^{2}+4\tau_{2}^{2}x^{2}+8\tau_{1}\tau_{3}x^{2}+8\tau_{2}\tau_{3}x^{2}\\
 &  & +4\tau_{3}^{2}x^{2}+1600\tau_{1}y-8t_{3}^{2}\tau_{1}y+1600\tau_{2}y+16t_{3}^{2}\tau_{2}y-8\tau_{1}^{2}\tau_{2}y-8\tau_{1}\tau_{2}^{2}y\\
 &  & +1600\tau_{3}y-8\tau_{1}^{2}\tau_{3}y+16\tau_{2}^{2}\tau_{3}y-8\tau_{1}\tau_{3}^{2}y+16\tau_{2}\tau_{3}^{2}y+16t_{3}^{2}y^{2}+16\tau_{1}^{2}y^{2}\\
 &  & +32\tau_{1}\tau_{2}y^{2}+16\tau_{2}^{2}y^{2}+32\tau_{1}\tau_{3}y^{2}+32\tau_{2}\tau_{3}y^{2}+16\tau_{3}^{2}y^{2}\\
 &  & +t_{1}^{2}(t_{2}^{2}+t_{3}^{2}+\tau_{2}^{2}+2\tau_{2}\tau_{3}+\tau_{3}^{2}+4x^{2}-8\tau_{2}y-8\tau_{3}y+16y^{2})\\
 &  & +t_{2}^{2}(4t_{3}^{2}+\tau_{1}^{2}-4\tau_{1}(\tau_{3}+2y)+4(\tau_{3}^{2}+x^{2}+4\tau_{3}y+4y^{2})).
\end{eqnarray*}

We need to solve the stationary equations, i.e., the derivatives of
$V$ with respect to $t_{1},t_{2},t_{3},\tau_{1},\tau_{2},\tau_{3},x$
and $y$ equated to zero. We also need to add an additional equation
$1-z(t_{1}t_{2}t_{3}x)=0$ to ensure that none of the denominators
of the stationary equations are zero. Thus, in total there are $9$
equations in $9$ variables. This system only has isolated solutions%
\footnote{In \cite{Gray:2006gn}, this system is reported to have positive dimensional
components in its solution space. However, the denominator equation
was not included in the analysis there. Once we include the denominator
equation in the system, the combined system has no positive dimensional
components. Hence, there is no discrepancy here.%
}. The equations are quite complicated and we avoid writing all of
them down here. This system of equations is not only \textit{prohibitively
difficult} to be solved completely but also not tractable even using
the primary decomposition techniques (except that some information
about the solutions may be obtained if one further restricts the system
such as taking $y=0$)\cite{Gray:2006gn,Gray:2009fy}. In short, it
is not possible to handle this system in its full glory using the
available symbolic methods.

Now, let us move to the NPHC method. Firstly, since there are four
equations of degree $3$, another four equations of degree $4$ and
one equation of degree $5$, the CBB is $103680$. This system is
actually quite straightforward to solve using the NPHC method. The
HOM4PS2 package, for example, solves the full system in around $1$
hour on a regular desktop machine: there are $516$ total solutions
for this system, out of which there are only $12$ real solutions.
The solutions in the order $\{y,\tau_{1},\tau_{2},\tau_{3},t_{2},t_{3},x,t_{1}\}$are:

\begin{eqnarray*}
\\
 &  & \{\{-3.3333333333335,1.3333333333308,3.333333333331,3.333333333336,\\
 &  & -6.6666666666666705,-6.666666666666667,-6.66666666666667,-2.6666666666603\},\\
 &  & \{-3.333333333334,1.3333333333288,3.333333333337,3.3333333333335,\\
 &  & 6.66666666667,6.666666666671,-6.666666666669,2.6666666666594\},\\
 &  & \{-3.3333333333344,1.3333333333324,3.33333333333,3.3333333333375,\\
 &  & -6.666666666668,-6.666666666666,6.666666666667,-2.6666666666634\},\\
 &  & \{-3.3333333333264,1.3333333333488,3.3333333333286,3.333333333322,\\
 &  & 6.666666666659,6.66666666666,6.666666666663,2.6666666666807\},\\
 &  & \{3.333333333338,-1.3333333333228,-3.333333333334,-3.333333333342,\\
 &  & 6.666666666674,6.666666666669,6.6666666666705,2.6666666666554\},\\
 &  & \{3.3333333333286,-1.3333333333406,-3.333333333338,-3.3333333333215,\\
 &  & 6.666666666663,6.666666666668,-6.666666666665,2.6666666666714\},\\
 &  & \{3.3333333333313,-1.3333333333337,-3.3333333333366,-3.3333333333277,\\
 &  & -6.666666666667,-6.6666666666705,6.666666666668,-2.666666666663\},\\
 &  & \{3.3333333333313,-1.3333333333341,-3.3333333333326,-3.333333333335,\\
 &  & -6.6666666666705,-6.666666666671,-6.666666666666,-2.6666666666634\},\\
 &  & \{0.,0.,0.,0.,7.453559924993,7.4535599249993,-7.453559925,2.9814239699997\},\\
 &  & \{0.,0.,0.,0.,7.453559924999,7.4535599249993,7.453559925,2.9814239699997\},\\
 &  & \{0.,0.,0.,0.,-7.4535599249992,-7.453559925,-7.4535599249992,-2.9814239699997\},\\
 &  & \{0.,0.,0.,0.,-7.4535599249993,-7.4535599249992,7.4535599249,-2.9814239699\}\}.
\end{eqnarray*}

It is easy to recognize that some of the numbers in the above list
of solutions are rational numbers, e.g., $3.33333\approxeq\frac{10}{3}$.
We can now easily compute the eigenvalues of the Hessian of the potential
and other related quantities of all these solutions and hence classify
the vacua in terms of physics. However, again we refrain from discussing
the interesting physics of these solutions here. The full analysis
will be published elsewhere.

\section{Comparison between Gröbner basis techniques and the NPHC method}

Here, we compare the two different methods. Firstly, the Gröbner basis
techniques solve the system symbolically. This is immensely significant
since one then has a \textit{proof} for the results and/or the results
in closed form. There is caveat here however: if the univariate equation
in a Gröbner basis is of degree $5$ or higher, then the Abel-Ruffini
theorem prevents us from solving it exactly, in general, at least
in terms of the radicals of its coefficients (this does not mean that
the univariate equation cannot be solved exactly at all). In such
a situation, one may end up solving this equation numerically and
hence the above mentioned feature of the symbolic method no longer
applies. The NPHC method is a numerical method. That said, the method
by construction gives \textit{all }of the isolated solutions for the
system known to have only isolated solutions, up to a numerical precision.
The solutions then can be refined to within an arbitrary precision
up to the machine precision by the Newton's corrector method or otherwise.
Moreover, using the alphaCertified method, we can certify if the real
non-singular solutions obtained by the above packages are actually
the real non-singular solutions of the system independent of the numerical
precision used during the computation. Hence, though the solutions
cannot be obtained in a closed form using the NPHC method, the solutions
are as good as exact solutions for all the practical purposes.

We should emphasis here that using the methods presented in \cite{Gray:2006gn,Gray:2008zs,Gray:2007yq}
one can learn quite a lot about a system without having to necessarily
obtain its solutions. In particular, one can use the so-called primary
decomposition of the ideal (though making use of the Gröbner basis
technique only) to obtain information such as the dimensionality of
the solution space, number of isolated real roots, etc. This is indeed
a clever way to resolve the above mentioned issue up to a certain
level. However, here, the next difficulty comes in the form of algorithmic
complexity. The BA is known to suffer from exponential space complexity,
which roughly means that the memory (Random Access Memory) required
by the machine blows up exponentially with increasing number of polynomials,
variables, monomials and/or degree of the polynomials involved in
the system. Hence, even the computation for the primary decomposition
may not finish for large sized systems, whereas the NPHC method is
strikingly different from the Gröbner basis techniques in that the
algorithm for the former suffers from no known major complexities.
Hence one can in principle find all solutions of bigger systems.

The BA is a highly sequential algorithm, i.e., each step in the algorithm
requires knowledge of the previous one. Thus, although recently there
are certain parts of the BA which have been parallelized, in general,
it is extremely difficult to parallelize the algorithm. On the other
hand, in the NPHC method, the path tracking is \textit{embarrassingly
parallelizable}, because each start solution can be tracked completely
independently of the others. This feature along with the rapid progress
towards the improvements of the algorithms makes the NPHC well suited
for a large class of physical problems arising not only in string
phenomenology but in condensed matter theory, lattice QCD, etc.

The BA is mainly defined for systems with rational coefficients, while
in real life applications, the systems may have real coefficients.
The NPHC method being a purely numerical method by default incorporates
floating point coefficients as well.

In conclusion, both the Gröbner basis techniques and the NPHC have
advantages and disadvantages. However, for practical purposes, the
NPHC method is a far more efficient and promising method for realistic
systems.

\section{Frequently Asked Questions}

In this Section, we collect the frequently asked questions and their
answers:
\begin{enumerate}
\item What does the NPHC method tell us about systems which do not have
any solutions?

As mentioned above, the NPHC method by construction (in conjunction
with the $\gamma$-trick), gives all real and complex solutions of
a system of multivariate polynomial equations that is known to have
only isolated solutions. Hence, we are always sure that we have got
all the solutions numerically. This statement is true for all cases
such as when the system has no complex solutions and/or no real solutions,
or no solution at all. One possible issue, as mentioned above, regards
the classification of the real solutions independent of the tolerance
used. This can be resolved by using, for example, the alphaCertified
algorithm which certifies when a solution is a real.

\item For many practical problems, only real solutions are required. Thus,
when implementing the NPHC, a huge amount of computational effort
is wasted in getting the other types of solutions. Would it not be
helpful to track only the real solutions?

It would be much more useful if there was a way of getting only real
solutions. However, for a number of technical reasons nicely discussed
in Ref.~\cite{SW:95}, a path tracker does not know in advance if
a given start solution will end up being a real solution of the original
system. Moreover, one can wonder if a root count exists only for the
real solutions of a system. This would involve obtaining a corresponding
fundamental theorem of algebra on the real space for the multivariate
case. This, however, has yet to be achieved. Hence, the best way for
now is to track all complex (including real paths) solutions and then
filter out the real solutions.

\item Can the NPHC method be used as a global or local minimization method?

Absolutely. Most of the conventional methods used to minimize a function
are based on the Newton-Raphson method, where a start solution is
guessed and is then refined by successive iterations in the direction
of the minima. By performing this algorithm several times on the functions,
one can obtain many minima of the given potential. Recently, more
efficient methods such as the basin-hopping method are available for
local minimization \cite{Wales:04}. However, we are never sure if
we have got all the minima from any of these methods. For the global
minimization, we may use more efficient methods such as Simulated
Annealing, Genetic Algorithm, etc. However, these methods are known
to fail for larger systems since it can easily get trapped at a local
minimum when trying to find the global minimum. Thus, in addition
to the usual error from the numerical precision of the machine, there
can be an error of an \textit{unknown} order (i.e., we do not know
if the found one is the global minimum!). But if the function has
a polynomial-like non-linearity, in theory the NPHC method can give
all the minima since it obtains all the stationary points. So it solves
the local minimization problem. Moreover, it is then easy to identify
\textit{the} global minimum out of the minima and hence we are sure
that the found one is actually the global minimum.

\item As in the example system in this paper, for many systems the number
of actual solutions may be well below the CBB. Is there any remedy
for this issue?

The main reason why the number of actual solutions is less than the
CBB for many systems is that the CBB does not take the sparsity (i.e.,
very few monomials in each polynomial in the system) of the system
into account. There is indeed a tighter upper bound on the number
of complex solutions, called the Bernstein-Khovanskii-Kushnirenko
(BKK) count \cite{SW:95,Li:2003}, which takes this sparsity into
account and thus in most cases is much lower than the CBB. In many
cases, it is in fact equal to the number of solutions. The BKK bound
can thus save a lot of computation time since the number of paths
to be tracked is less than the CBB. The details on the BKK count relating
to string phenomenology problems will be published elsewhere.

\item Are there any alternative/supplementary numerical methods?

There are not many methods to find the stationary points of a multivariate
function around, compared to the number of methods to find minima.
One of the methods that can find stationary points is the Gradient-minimization
method which finds all the minima of an auxiliary function $E=|\nabla V|^{2}$
whose minima are the stationary points of $V$ provided we further
restrict $E$ to be zero \cite{PhysRevLett.85.5360}. One can find
many minima of $E$ using some conventional minimization method such
as the Conjugate Gradient method or the Simulated Annealing method.
However, it is known that as the system size increases, the number
of minima of $E$ that are not the minima of $V$, i.e. $E>0$, increases
rapidly, making the method inefficient \cite{2002JChPh.116.3777D}.
Another method is the Newton-Raphson method (and its sophisticated
variants) \cite{2002JChPh.116.3777D,2003JChPh.11912409W,2002PhRvL..88e5502G,Wales:04}.
There, an initial guess is refined iteratively to a given precision.
It should be emphasized, however, that no matter how many different
random initial guesses are fed into the algorithm, we can never be
sure to get all the solutions in the end, unlike the NPHC method.
However, these two methods can be supplementary methods for bigger
systems to get an idea on what to expect there.

\item This paper mainly deals with the potentials having polynomial like
non-linearity which may be usual in the perturbation limit. What about
the fully non-perturbative potentials?

The most interesting application for this method would be in the non-perturbative
regime, certainly. This question can be stated in different words:
is it possible to translate the stationary equations for the non-perturbative
potential (i.e., the potentials which have logarithm and exponential
terms), and if so, how? Once we can translate the equations in the
polynomial form, we can again use the NPHC method as before. The answer
is already available in \cite{Gray:2007yq}. In this work, Gray et
al. have already prescribed how to translate the corresponding equations
arising in the non-perturbative regime which usually involve logarithms
and/or exponentials, by using dummy variables. After that, we can
solve the system using algebraic geometry methods, such as the Gröbner
basis, or for more complicated cases, the NPHC method presented in
this paper. Once we have all the solutions, we can extract the solutions
in terms of the original variables which were logarithms and/or exponentials
of the fields. This trick makes all the algebraic geometry methods,
not only the NPHC method, applicable to finding the vacua of the potentials
in the non-perturbative regime.

\item This method assumes that one knows that the system under consideration
has only isolated solutions. But, in general, one may not know if
a given system has only isolated solutions or it contains some positive
dimensional components. In that case, don't we need to rely on the
Gröbner basis techniques only, at least to check the dimension of
the system?

Firstly, in most systems available in the example-suit of the Stringvacua
package, once we add the constraint equation (i.e., the denominators
are never zero), they usually turn out to have only isolated solutions.
Thus, there are way too many interesting systems in string phenomenology
which only have isolated solutions. Of course, there may be many more
systems which would have positive dimensional solution-components.
To solve such systems, there is a recently developed generalization
of the numerical homotopy continuation method, called the Numerical
Algebraic Geometry method. This method finds out each of the positive
dimensional solution-components with its dimensionality. This method
is also \textit{embarrassingly parallelizable} and hence goes far
beyond the reach of the Gröbner basis methods. The details of this
method are much more involved and beyond the scope of the present
article. But, in short, to find out the dimensionality of the system
we do not necessarily need to rely on the Gröbner basis methods. The
details of this method with applications will be published elsewhere.

\end{enumerate}

\section{Summary}

In this paper, we have reviewed a novel method, called the numerical
polynomial homotopy continuation (NPHC) method, which can find all
the string vacua of a given potential. It does not suffer from any
major algorithmic complexities compared to the existing symbolic algebra
methods based on the Gröbner basis techniques, which are known to
suffer from exponential space complexity. Moreover, the NPHC method
is \textit{embarrassingly parallelizable, }making it a very efficient
alternative to the existing symbolic algebra methods\textit{. }As
an example, we studied a toy model and, using the NPHC method, found
all the vacua within less than a minute using a regular desktop machine.
Note that this system with the irrational coefficients is already
a difficult task using the Gröbner basis techniques. In addition to
that, using the NPHC method, with just about an hour of computation
on a regular desktop machine, we found all vacua of an M theory model
compactified on the coset $\frac{SU(3)\times U(1)}{U(1)\times U(1)}$,
which has an $SU(3)$ structure. This system was reported to be a
prohibitively difficult problem using the symbolic method. Thus, we
have already shown how efficiently the NPHC method can solve the problems
that are yet far beyond the reach of the traditional symbolic methods.
We also emphasis that using the procedure prescribed in \cite{Gray:2007yq}
to translate the stationary equations arising in the non-perturbative
regime, by replacing logarithm and exponential terms of the field
variables by dummy variables, into the polynomial form, we can use
the NPHC method to find the vacua for the non-perturbative potentials
as well. It is this application of the method which makes it quite
promising. With the help of the NPHC method it is thus hoped that
we can go far beyond the reach of the existing methods and study realistic
models very efficiently.

\section{Acknowledgment}

DM was supported by the U.S. Department of Energy grant under contract
no. DE-FG02-85ER40237 and Science Foundation Ireland grant 08/RFP/PHY1462.
DM would like to thank James Gray, Yang-Hui He and Anton Ilderton
for encouraging and helping him throughout in this work.

\bibliographystyle{plain}
\bibliography{bibliography}

\end{document}